\documentclass[
  aps,
  prl,
  reprint,
  longbibliography,
  nofootinbib,
  floatfix,
  amsmath,
  amssymb
]{revtex4-2}

\usepackage{graphicx}
\usepackage[version=3]{mhchem}
\usepackage{xspace}
\usepackage[hidelinks]{hyperref}

\newcommand{\ThCaF}{\ce{^229Th}{:}\ce{CaF2}\xspace}
\newcommand{\thor}{\ensuremath{^{229}\mathrm{Th}}\xspace}

\newcommand{\Vzz}{V_{zz}}
\newcommand{\rQ}{r_Q}
\newcommand{\gammaU}{\gamma_{3}}

\newcommand{\nuS}{\nu_{3}}

\newcommand{\fwhmouter}{\mathrm{FWHM}_{\text{outer}}}
\usepackage[dvipsnames]{xcolor}

\begin{document}

\title{Disorder-induced nuclear-transition doublet at nominally cubic \texorpdfstring{$^{229}$Th}{229Th} sites}

\author{A. Derevianko}
\affiliation{Department of Physics, University of Nevada, Reno, Nevada 89557, USA}

\date{\today}

\begin{abstract}
{The \ce{^229Th} nuclear clock transition is inhomogeneously broadened even at
sites of nominally cubic symmetry, where the electric-field gradient of the
ideal lattice vanishes.  We show that dilute static point-defect disorder
produces a doublet with a zero at the quadrupole-free frequency.  The doublet
is symmetric in the isotropic approximation; size misfit point defects introduce a small
lobe asymmetry without removing the zero.  Charged and size-misfit defects
share inverse-cube scaling and an approximately common normalized profile,
whose width separates into a host-controlled background and a term linear in
the \ce{^229Th} density.  In a heavily doped \ThCaF{} crystal
[Nature~657, 626 (2026)], the thorium-linked contribution is comparable to the
measured $1.1~\mathrm{MHz}$ width; in a lightly doped crystal
[arXiv:2606.08870], the $0.9~\mathrm{MHz}$ width bounds the host background.
Neither spectrum shows the predicted dip,  which points to the presence of extended defects in both crystals.
If the predicted dip is resolved, it can provide a nuclear-clock discriminator referenced to the quadrupole-free nuclear transition frequency.}
\end{abstract}

\maketitle

The $8.4\,\mathrm{eV}$ isomeric transition in \thor{} can be driven at
$148.4\,\mathrm{nm}$ and interrogated either in trapped ions or in a
solid-state host~\cite{PieTam03,CamRadKuz12,Rellergert2010,
Derevianko2026ColloquiumNuclearClocks}.  A doped crystal offers a macroscopic
ensemble of quantum oscillators in a compact package, at the cost of material-dependent
shifts of the transition frequency (isomer
shifts)~\cite{perera2025-isomer-shift-Th229}, relaxation
channels~\cite{Schaden2024-Th-quenching,Morgan2025_internal_conversion}, and
inhomogeneous broadening~\cite{Higgins2025-Temperature-dependence-ThCaF2,
Ooi2026FrequencyReproducibility}.  Laser excitation has been demonstrated in
\ThCaF~\cite{Tiedau2024-caf2}, in \ce{^229Th}{:}\ce{LiSrAlF6}~\cite{Elwell2024-lisaf},
and in \ce{ThF4} and \ce{ThO2} films~\cite{Zhang2024-ThF4,Elwell2025-ThO2}.

In \ThCaF{}, laser M\"ossbauer spectroscopy resolved four thorium populations
distinguished by their electric-field gradients
(EFGs)~\cite{Hiraki2025-caf2-site-resolved}.  The EFG is the second derivative
of the electrostatic potential at the nucleus; it couples to the nuclear
quadrupole moment and splits otherwise degenerate nuclear spin manifolds.  One
of the four populations showed no resolved quadrupole structure and a fitted
EFG near zero, consistent with a \ce{Th^4+} ion substituting for \ce{Ca^2+} at
a site of full octahedral symmetry, where the EFG of the ideal lattice vanishes
identically~\cite{CohenReif1957-QuadrupoleEfectsNMR}.  Following the
 literature we call it the \emph{O center}, derived from the
octahedral point group $O_h$ (the symmetry group of a cube, including
inversion).

The O center can be the natural nuclear clock candidate: with no static gradient there is
no quadrupole splitting to control.  Continuous-wave (CW) measurements~\cite{Morawetz2026CwAbsorption,Huang2026NuclearClock}  in \ThCaF have since shown that cubic symmetry does not deliver a narrow line. Neither the microscopic origin of the width nor the shape of the underlying distribution
has been established. Here we address this gap.

\begin{figure}[th]
\includegraphics[width=0.85\columnwidth]{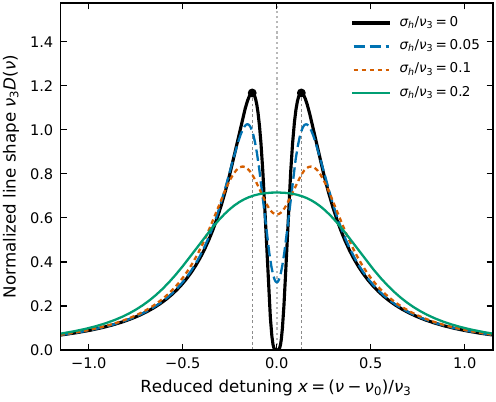}
\caption{\label{fig:lineshape}Calculated \thor{} nuclear line shape $D(\nu)$ at
a nominally cubic site before (black solid
curve) and after convolution with normalized Gaussian response functions of
standard deviation $\sigma_h/\nuS=0.05$ (blue dashed), $0.10$ (orange dotted),
and $0.20$ (green solid); the black curve has $\sigma_h=0$.  The horizontal
coordinate is the reduced detuning $x=(\nu-\nu_0)/\nuS$, where $\nu$ is the
laser frequency, $\nu_0$ the quadrupole-free transition frequency, and
$\nuS=2Q_g\gammaU/h$ the EFG frequency scale of Eq.~\eqref{eq:nuscale}; the
vertical coordinate is the dimensionless profile $\nuS D(\nu)$.  Steep inner walls of a resolved dip can be used for nuclear clock operation.}
\end{figure}

The \thor{} ground and isomeric nuclear states have
spins $I_g=5/2$ and $I_m=3/2$ and spectroscopic electric quadrupole moments
$Q_g$ and $Q_m$, whose ratio is
$\rQ\equiv Q_m/Q_g=0.57003$~\cite{Zhang2024-Th229Comb}; we use
$Q_g=3.11\,e\,\mathrm b$~\cite{PorSafKoz2021-Th3plusHFS}. EFG tensor $\mathbf V$ is traceless and symmetric.
In its principal-axis frame (PAF), quadrupole coupling reads
\begin{equation}
H_Q=\frac{Q\Vzz}{4I(2I-1)}
\left[3I_z^2-I(I+1)+\eta\left(I_x^2-I_y^2\right)\right],
\label{eq:Hq}
\end{equation}
where $I_k$ are the nuclear spin components in PAF. EFG's principal components are labeled
by the convention $|\Vzz|\ge|V_{yy}|\ge|V_{xx}|$, and the  asymmetry parameter $\eta=\frac{V_{xx}-V_{yy}}{\Vzz}$,
$0\le\eta\le1$ measures departure from axial
symmetry. Time reversal pairs the eigenstates of Eq.~\eqref{eq:Hq} into Kramers
doublets, two in the isomeric and three in the ground manifolds, so
the quadrupole coupling nominally results in six unique nuclear transitions (a sextet). For $\mathbf{V} \equiv 0$, the sextet collapses into a single line.

{Here we show that (i) dilute static EFG disorder at a cubic site produces
a \emph{doublet} with a zero at the EFG-free frequency $\nu_0$; its symmetric
analytic approximation is shown in Fig.~\ref{fig:lineshape}, while size misfit
can make the lobes slightly unequal; (ii) charged and size-misfit defects share
inverse-cube stable-law scaling and an approximately common normalized line
shape; and (iii) because charge neutrality constrains only the \emph{signed}
charge sum, the linewidth separates into a host-controlled term and a
thorium-linked term, which the two published spectra constrain separately.
Technical details can be found in SM~\cite{SupplementalMaterial}.}\nocite{Safronova2018ThRadii,Shirley1964-isomer-rmp,BornHuang1954,Andeen1971,Czjzek1982EFGAmorphousMaterials,DerPerKro2026-EFGs,Gale1997GULP,Shi2005-CaF2-Fcenters-abinitio,Eshelby1954PointImperfections,Tinkham1964GroupTheory,Hill1952ElasticAggregate,Holtsmark1919Verbreiterung,Chandrasekhar1943StochasticProblems,KittelAbrahams1953DipolarBroadening}

\ce{CaF2} is an  ionic band insulator. Qualitatively, in pristine \ce{CaF2}, each \ce{Ca^2+} ion is surrounded by eight \ce{F^-} ions located at the corners of a cube (first coordination shell).
Although the individual \ce{F^-} ions make large contributions to the EFG at the Ca site, cubic symmetry causes their sum to vanish. Even small lattice distortions can lift this cancellation. When closed-shell \ce{Th^4+} substitutes for \ce{Ca^2+}, all eight neighboring fluorine ions are drawn inward symmetrically and  the EFG remains zero. However, the symmetry is broken by the need to compensate the  \ce{Th^4+}-\ce{Ca^2+}  charge mismatch by introducing either two fluorine interstitials $\mathrm F_i'$ or by one
calcium vacancy $\mathrm V_{\mathrm{Ca}}''$~\cite{kroger1974-book,Dessovic2014,
Pimon2022-ab-initio-CaF2-heavy,Rehn2026DefectAssignment}.\footnote{We name defects in Kr\"oger--Vink notation~{\cite{KrogerVink1956,kroger1974-book}}, in which the
main symbol is the species, the subscript is the site it occupies ($i$ for an
interstitial, $\mathrm V$ for a vacancy), a superscript dot denotes one unit of
positive effective charge relative to the ideal lattice, and a superscript
prime one unit of negative effective charge.}
When the two $\mathrm{F}_{i}^{\prime}$ occupy interstitial sites near or inside the first coordination shell, they displace the host \ce{F^-} ions, strongly break the local symmetry, and generate large EFGs. Such configurations are associated with some of the spectral sub-populations, e.g. D centers, observed experimentally.

As the \ce{F^-} interstitials occupy progressively more distant coordination shells, the induced lattice distortions and E-fields weaken and the EFG at the \thor{} site decreases. Random spatial configurations of these remote interstitials  produce a distribution of EFGs, leading to site-dependent inhomogeneous broadening of the nuclear transition. The EFG at the probe \thor{} nucleus receives contributions from both the direct, crystal-mediated electrostatic effects of the interstitials and the lattice distortions that they induce in the ionic host. Real crystals, beyond the desired Th defects, contain various impurities and point defects that also contribute to EFGs and inhomogeneous broadening. Below we consider microscopic EFG contributions of the charged and size misfit point defects.

Let $\mathbf r$ be the position of the probe \thor{} nucleus
measured from the defect.  A defect carrying a charge $q$ relative
to the host produces the EFG tensor
\begin{equation}
V_{ij}^{(q)}=q\, {A_{ijkl}}\, {K_{kl}}(\mathbf r),
\qquad
K_{ij}(\mathbf r)=\frac{3r_ir_j-r^2\delta_{ij}}{r^5},
\label{eq:chargeefg}
\end{equation}
where $K_{ij}$ is the point-charge EFG kernel, repeated indices are
summed over, and  $A_{ijkl}$ is the EFG response of the host,
collecting its lattice polarization and displacement together with the
electronic antishielding of the \ce{Th^4+} ion.

A {\em neutral} defect whose size differs from that of the site it occupies acts instead
through elastic strain.  At long range it behaves as a center of dilatation of
far-field volume $\Delta V$ with radial displacement field
$\mathbf u(\mathbf r)=(\Delta V/4\pi)\hat{\mathbf r}/r^2$ causing the strain
tensor $\varepsilon_{ij}=\tfrac12(\partial_iu_j+\partial_ju_i)=-\frac{\Delta V}{4\pi}K_{ij}(\mathbf r)$~\cite{Eshelby1956ContinuumTheory}. Then the EFG
\begin{equation}
{V^{(s)}_{ij}= -(1-\gamma_\infty)\frac{\Delta V}{4\pi} G_{ijkl}K_{kl}(\mathbf r)\,,}
\label{eq:misfitstrain}
\end{equation}
where $G_{ijkl}$ is (unshielded) gradient-elastic tensor~\cite{CohenReif1957-QuadrupoleEfectsNMR,Shulman1957GradientElastic}.
{The two defect types, charged and size-misfit, depend on the same
inverse-cube source tensor and differ through the host response acting on it.
We replace $A_{ijkl}$ and $G_{ijkl}$ by isotropic-equivalent magnitudes
$A_{\mathrm{eff}}$ and $G_{\mathrm{eff}}$, which preserve the mean EFG magnitude
averaged over orientations of $K_{ij}$.  For the separately tested response
channels this changes the outer FWHM by less than $1\%$, despite the strong
response anisotropy.  Within this scalar approximation the prefactors are
$A_{\mathrm{eff}}q$ and $-(1-\gamma_\infty)G_{\mathrm{eff}}\Delta V/4\pi$, where
$\gamma_\infty=-177.5$ is the Sternheimer antishielding factor of the
\ce{Th^4+} ion~\cite{FeiJoh69}.  Both channels share inverse-cube stable-law
scaling; their normalized profiles coincide within the scalar-response and
$\mathrm{SO}(5)$-isotropic approximations.  Deviations in the inner profile and
lobe symmetry, and the separate $2\%$ assessment of the combined scalar
charge--misfit response in the reinforcing regime, are given in
SM~\cite{SupplementalMaterial}.  Strain disorder of this kind was shown to
produce Lorentzian statistics in nuclear magnetic
resonance~\cite{Stoneham1969ShapesInhomogeneouslyBroadened}.}

Cubic symmetry restricts each EFG response to two independent components,
acting separately on the $E_g$ and $T_{2g}$ parts of a symmetric traceless
tensor, i.e. in the cube frame, its diagonal and off-diagonal components.
The same symmetry forces a second-rank response such
as the dielectric tensor to be a scalar, but leaves a fourth-rank response
anisotropic.  \ce{CaF2} displays both behaviors optically: a single refractive
index, yet three independent photoelastic constants~{\cite{VeerabhadraRao1970PhotoelasticCaF2}} and an intrinsic
spatial-dispersion birefringence~\cite{Burnett2001IntrinsicBirefringence}.  A scalar dielectric constant
therefore cannot represent the response to a field \emph{gradient}.
Evaluating the two components for \ce{CaF2} with the shell-model
polarizabilities~\cite{CatlowNorgett1973}, we obtain the gradient-elastic components
$G_{E_g}=-13.3$ and $G_{T_{2g}}=+18.9~\mathrm{V/\text{\AA}^2}$, with
isotropic-equivalent magnitude $G_{\mathrm{eff}}=16.8~\mathrm{V/\text{\AA}^2}$.
The charge channel is far more anisotropic: $A_{E_g}/A_{T_{2g}}=+0.03$ to
$+0.09$ for the electronic polarization response and $-0.22$ once the ions are allowed to
relax, bracketing $A_{\mathrm{eff}}=1.8$--$2.3\times10^{2}$; we adopt the
conservative value $A_{\mathrm{eff}}=180$. This corresponds to effective relative dielectric constant $\approx 1$.   Then a far-field volume of $10.9~\text{\AA}^3$
contributes as much to the disorder as one elementary charge $e>0$.  Thereby,
the equivalent  \emph{ ``elastic'' charge} of a size-misfit defect is
\begin{equation}
q_{\mathrm{eq}}(\Delta V)=
{\frac{(1-\gamma_\infty)G_{\mathrm{eff}}}{4\pi A_{\mathrm{eff}}}\,\Delta V=}
0.092\,e\,
\left(\frac{\Delta V}{1~\text{\AA}^3}\right).
\label{eq:qeq}
\end{equation}
See SM~\cite{SupplementalMaterial} for our estimated $\Delta V$ values for  representative \ce{CaF2} defects.

\textit{EFG statistics.---}The EFG tensor $\mathbf V$, with laboratory
components $V_{ij}$, is symmetric and traceless and therefore has five
independent components.  We represent $\mathbf V$ by a five-component real vector
$\mathbf U=(U_1,\ldots,U_5)$ in the  irreducible basis~\cite{Czjzek1981AtomicCoordinationEFG,LeCaer2010ExtendedCzjzekEFG}, in
which a physical rotation $R\in\mathrm{SO}(3)$ of the laboratory axes acts
as an orthogonal transformation of $\mathbf U$.  Then its  length
is  invariant under physical rotations (with $\mathcal M(\eta)=\sqrt{1+\eta^2/3}$)
\begin{equation}
\lVert\mathbf U\rVert = \left(\sum_{k=1}^5U_k^2\right)^{1/2}
=\left[\frac{\operatorname{tr}(\mathbf V^2)}{6}\right]^{1/2}
=\frac{|\Vzz|}{2}\,\mathcal M(\eta)\,.
\label{eq:invariant}
\end{equation}

To find the joint probability density $P(\mathbf U)$ of the
five EFG components, we compute characteristic function $\Phi(\mathbf t)$;
these two are related by a 5D Fourier transformation with five-vector $\mathbf t$ being the Fourier conjugate to $\mathbf U$.
Let $\alpha$ label a defect species and $\ell$ one of its allowed lattice sites,
let $p_{\alpha\ell}$ be the probability that site $\ell$ is occupied by species
$\alpha$, and let $\mathbf u_{\alpha\ell}$ be the EFG five-vector that this single
defect contributes at the probe.  For statistically independent site
occupations the characteristic function factorizes,
\begin{equation}
\Phi(\mathbf t)=\big\langle e^{i\mathbf t\cdot\mathbf U}\big\rangle
=\prod_{\alpha,\ell}
\left[1-p_{\alpha\ell}+p_{\alpha\ell}\,
e^{i\mathbf t\cdot\mathbf u_{\alpha\ell}}\right],
\label{eq:characteristic}
\end{equation}
where
$\langle\cdots\rangle$
denotes an ensemble average.

In the dilute continuum limit relevant to experiments, with the $\mathrm{SO}(5)$
isotropy approximation~\cite{Stockmann1981UnscreenedPointChargeEFG}, in which the
probability depends only on $\lVert\mathbf U\rVert$ and not on the 5D-space  direction of
$\mathbf U$, the $r^{-3}$ kernel gives
$\ln\Phi=-\gammaU\lVert\mathbf t\rVert$. Joint distribution $P(\mathbf U)$ of the five EFG components is then
\begin{equation}
P(\mathbf U)=\frac{2\gammaU}
{\pi^3\left(\gammaU^2+\lVert\mathbf U\rVert^2\right)^3}, \label{eq:cauchy}
\end{equation}
where
$\gammaU$ is the material-dependent EFG width; its subscript records the $r^{-3}$ law shared by the charged and
size-misfit universality class.  We find numerically that the $\mathrm{SO}(5)$ isotropy
{approximation~\cite{Stockmann1981UnscreenedPointChargeEFG} gives an outer
FWHM and peak separation larger than direct sampling with the true
point-charge kernel: the sampled values are $2.7\%$ and $2.0(5)\%$ below the
analytic values, respectively~\cite{SupplementalMaterial}.}

Eq.~\eqref{eq:cauchy} yields the probability density of the EFG magnitude $R=\lVert\mathbf U\rVert/\gammaU$ (5D sphere surface area is $8\pi^2 R^4/3$)
\begin{equation}
f(R)=\frac{16}{3\pi}\frac{R^4}{(1+R^2)^3}\,.
\label{eq:cauchy-fR}
\end{equation}
As $R\rightarrow 0$,  $f(R)$ vanishes as $R^4$: a small
EFG requires all five components to be small at once, and the {\em eigenvalue
repulsion} of a real symmetric matrix suppresses near-degenerate configurations.
{This quartic law does not rely on the $\mathrm{SO}(5)$ isotropy approximation.  For
any EFG distribution that is invariant under spatial rotations, the joint
probability density of the eigenvalues $\lambda_i$ of $\mathbf V$ has the
form~\cite{Rosenzweig1960,Guhr1997,Czjzek1981AtomicCoordinationEFG}
\begin{equation}
dP\propto\delta\Big(\sum_i\lambda_i\Big)
F\Big(\sum_i\lambda_i^2,\prod_i\lambda_i\Big)
\prod_{i<j}|\lambda_i-\lambda_j|\prod_id\lambda_i ,
\label{eq:eigmeasure}
\end{equation}
where the  factor $\prod_{i<j}|\lambda_i-\lambda_j|$ expresses the
eigenvalue repulsion.  Together with the trace constraint it gives
$f(R)\propto R^4$ as $R\to0$ whenever $F$ has a finite, nonzero limit at the
origin, i.e., whenever $P(\mathbf U)$ is bounded and nonzero at
$\mathbf U=0$.  When $F$ depends on $\sum_i\lambda_i^2$ only, as in the
$\mathrm{SO}(5)$ approximation,} the asymmetry $\eta$ is statistically independent of $R$ and distributed as
$p_\eta(\eta)=3\eta(1-\eta^2/9)(1+\eta^2/3)^{-5/2}$; it is nearly uniform with
 $\langle\eta\rangle=0.6098$~\cite{LeCaer2010ExtendedCzjzekEFG}.

The product in Eq.~\eqref{eq:characteristic} makes $\ln (\Phi)$ additive over
defect species, so  all the
material physics enters through the sum,
\begin{equation}
\gammaU=\frac{4\pi^{2}}{9\sqrt3}A_{\mathrm{eff}}
\sum_\alpha n_\alpha\bigl|\tilde q_\alpha\bigr|,
\qquad
\tilde q_\alpha\equiv q_\alpha-q_{\mathrm{eq}}(\Delta V_\alpha),
\label{eq:gammascale}
\end{equation}
where $n_\alpha$ is the number density of species $\alpha$, $q_\alpha$ its
charge relative to the host lattice,
and $q_{\mathrm{eq}}(\Delta V)$ is the  ``elastic''
charge~\eqref{eq:qeq}.  We call $\tilde q_\alpha$ the \emph{EFG source charge}: the far-field
EFG of one such defect is $A_{\mathrm{eff}}\tilde q_\alpha K_{ij}(\mathbf r)$.
Its values are evaluated in the SM~\cite{SupplementalMaterial}: for isolated
defects $|\tilde q_\alpha|$ ranges from $1.04\,e$ for
$\mathrm V_{\mathrm F}^{\bullet}$ to $2.74\,e$ for $\mathrm V_{\mathrm{Ca}}''$; a charge-neutral bound Th complex retains
$|\tilde q_\alpha|=0.72$--$1.05\,e$ through its size misfit.

\textit{The doublet.---}
A laser drives the \thor{} sextet by coupling to its magnetic-dipole (M1) transition-dependent moments.
Let $s=\pm1$ label the isomeric ($I_m=3/2$) and $n=0,1,2$ the ground-state ($I_g=5/2$) Kramers doublets.
Branch $(s,n)$ is EFG-shifted
from $\nu_0$ by
\begin{equation}
\Delta\nu_{sn}=\pm\nuS\,a_{sn}(\eta)\,R,
\,\,
a_{sn}(\eta)=\frac{s\rQ}{4}-\frac{\mu_n(\eta)}{40\,\mathcal M(\eta)},
\label{eq:branches}
\end{equation}
the sign being that of $\Vzz$.  Here $a_{sn}$ is the dimensionless branch
coefficient, the $\mu_n(\eta)$ are the three analytically-known roots that give the
ground-state energies $E_{g,n}=Q_g\Vzz\mu_n(\eta)/40$
\cite{Derevianko2026ColloquiumNuclearClocks}. The EFG frequency scale is
\begin{equation}
\nuS=\frac{2Q_g\gammaU}{h}
=15.0~\mathrm{MHz}\,
\frac{\gammaU}{1~\mathrm{V/\text{\AA}^2}} \,.
\label{eq:nuscale}
\end{equation}
The orientation-averaged M1 transition strengths are
$W_{sn}(\eta)=1/6+s\mu_n(\eta)/[60\,\mathcal M(\eta)]$, normalized so that
$\sum_{s,n}W_{sn}=1$.

Writing $\nu$ for
the laser frequency and averaging over $p_\eta(\eta)$  gives the  line shape ($\int D(\nu)d\nu=1$)
\begin{equation}
D(\nu)=\frac{1}{\nuS}\sum_{s,n}\int_0^1\! d\eta\,
\frac{p_\eta(\eta)\,W_{sn}(\eta)}{2|a_{sn}(\eta)|}\,
f\!\left(\frac{|\nu-\nu_0|}{\nuS|a_{sn}(\eta)|}\right).
\label{eq:lineshape}
\end{equation}
It has a closed form, see SM~\cite{SupplementalMaterial}.

Two features of Eq.~\eqref{eq:lineshape} are robust.  First, for the \thor{} Q-moment ratio $\rQ$  none of the
six coefficients $a_{sn}(\eta)$ vanishes. Then  every
branch  reaches $\nu_0$ only when the \emph{entire} EFG tensor vanishes,
so the $R^4$ suppression in Eq.~\eqref{eq:cauchy} transfers directly to the
spectrum,
\begin{equation}
D(\nu)\propto|\nu-\nu_0|^4
\quad(\nu\to\nu_0),
\qquad
D(\nu)\propto|\nu-\nu_0|^{-2}
\label{eq:quartic}
\end{equation}
in the wings, $|\nu-\nu_0|\gg\nuS$.  {The quartic law is asymptotic: the
anisotropic charge response, $A_{E_g}/A_{T_{2g}}\ll1$, compresses the two $E_g$
components, so over most of the dip the profile rises nearly as
$|\nu-\nu_0|^2$~\cite{SupplementalMaterial}.}
{Second, when $P(\mathbf U)=P(-\mathbf U)$, every realization has an
equally probable partner with the opposite frequency shift, and the profile
is symmetric about $\nu_0$, as in Fig.~\ref{fig:lineshape}.  This symmetry is exact for
charged defects once the charge neutrality is imposed.  Size misfit defects introduce minor lineshape asymmetry, see SM~\cite{SupplementalMaterial}.

Two spectroscopic precedents for a central dip rest on different physics.
At the  cubic Fe$^{2+}$ site in MgO, random strain lifts the orbital
degeneracy of the \emph{open-shell} Fe$^{2+}$ ion whose own $3d$ shell then produces an
EFG of strain-independent magnitude.  Its $^{57}$Fe M\"ossbauer spectrum
is a quadrupole doublet of fixed splitting, resolved only while electronic
relaxation is slow~\cite{Ham1967MossbauerMgO,LeiderPipkorn1968MgO,
Tjon1968MossbauerSpectraFluctuatingII,Chappert1970StressMgO}.  In amorphous
solids, whose sites are not cubic, Czjzek's random-coordination model predicts
the zero-EFG hole, but the M\"ossbauer spectra analyzed with it are too
poorly resolved to show the hole directly~\cite{Czjzek1981AtomicCoordinationEFG,
Czjzek1982EFGAmorphousMaterials}. The \thor O center
differs from both: \ce{Th^4+} is  closed-shell ($J=0$)  and has exactly zero EFG in the ideal
$O_h$ lattice, while remote defects generate EFGs with no pre-existing principal axis. To our knowledge
this closed-shell cubic-site mechanism remains unobserved.

Writing $\Delta\nu_{\mathrm{peak}}$ for the
separation of its two maxima and $\fwhmouter$ for the separation of the two
outer half-maximum points,
\begin{equation}
\Delta\nu_{\mathrm{peak}}=0.26131\,\nuS, \,
\fwhmouter=0.64666\,\nuS .
\label{eq:metrics}
\end{equation}
Note that $\fwhmouter$ is unconventional, because the
interval includes the central dip. The analytic lineshape is an approximate result; direct
Monte-Carlo sampling with the exact $K_{ij}$ kernel narrows the profile,
giving $\fwhmouter=0.629(4)\,\nuS$ {and $\Delta\nu_{\mathrm{peak}}=0.256(1)\,\nuS$}.  The quoted uncertainty {of $\fwhmouter$} combines the sampling noise with the
anisotropy of the two rank-four response tensors, which leaves $\fwhmouter$
unchanged but moves the maxima inward, to $\Delta\nu_{\mathrm{peak}}=0.240$--$0.251\,\nuS$,
and narrows the central dip~\cite{SupplementalMaterial}.

As to the nuclear clock, the steep inner walls of a resolved dip can provide a sharp frequency
discriminator.  For an even profile and balanced probe signals its zero
crossing lies at $\nu_0$; lobe asymmetry introduces a composition-dependent
offset that must be characterized.

EFG width $\gammaU$, Eq.~\eqref{eq:gammascale},
depends on $\sum_\alpha n_\alpha\bigl|\tilde q_\alpha\bigr|$, whereas macroscopic charge
neutrality requires $\sum_\alpha n_\alpha q_\alpha=0$. These are not equivalent even ignoring the size misfits as $|\tilde q_\alpha|  \rightarrow |q_\alpha|$ in this case.
The disorder can be separated into two populations with
different dependences on the \thor{} concentration.

The first is thorium-linked.  A \thor fraction $f_{\mathrm O}$
occupies O centers and is compensated remotely; the remaining $1-f_{\mathrm O}$
is bound into locally compensated
complexes~\cite{kroger1974-book,Dessovic2014,Pimon2022-ab-initio-CaF2-heavy,
Rehn2026DefectAssignment}.  Charge neutrality within the O-center population
requires either $2f_{\mathrm O}n_{\mathrm{Th}}$ interstitials $\mathrm F_i'$ or
$f_{\mathrm O}n_{\mathrm{Th}}$ vacancies $\mathrm V_{\mathrm{Ca}}''$, or their mix.  Within O centers
$\tilde q_{\mathrm O}=|\tilde q_{\mathrm{Th}_{\mathrm{Ca}}^{\bullet\bullet}}|
+2|\tilde q_{\mathrm F_i'}|=4.6$--$4.9\,e$ or
$|\tilde q_{\mathrm{Th}_{\mathrm{Ca}}^{\bullet\bullet}}|
+|\tilde q_{\mathrm V_{\mathrm{Ca}}''}|=4.5$--$4.9\,e$, equal within
uncertainty.  A bound complex enters only through its size misfit,
$|\tilde q_{\mathrm b}|=0.72$--$1.05\,e$~\cite{SupplementalMaterial}.
The thorium-linked part of Eq.~\eqref{eq:gammascale} is then
$\tilde q_{\mathrm{eff}}n_{\mathrm{Th}}$, with
\begin{equation}
\tilde q_{\mathrm{eff}}=f_{\mathrm O}\,\tilde q_{\mathrm O}+(1-f_{\mathrm O})\,|\tilde q_{\mathrm b}| .
\label{eq:qeffsplit}
\end{equation}
The second population is charge balanced within itself and independent
of $n_{\mathrm{Th}}$. We write its
contribution to Eq.~\eqref{eq:gammascale} as
$e\,n_{\mathrm{bg}}\equiv\sum_{\alpha\in\mathrm{bg}}n_\alpha|\tilde q_\alpha|$,
so that $n_{\mathrm{bg}}$ counts elementary EFG source charges rather than
defects.  A
dissociated anion Frenkel pair, $\mathrm V_{\mathrm F}^{\bullet}$ with an
uncorrelated $\mathrm F_i'$~\cite{Ure1957}, contributes $2.42\,e$, and a
Schottky-like set, $\mathrm V_{\mathrm{Ca}}''$ with two
$\mathrm V_{\mathrm F}^{\bullet}$, $4.82\,e$~\cite{SupplementalMaterial}.
Impurities, whose valence differs from that of the host ion they
replace,  with remote compensators~\cite{Corish1982,Bendall1984}
and neutral size-misfit species add to it. VUV laser, ionizing radiation, and the
$\alpha$-decay of \thor{} and its daughters continuously create and transform such
defects~\cite{Lindner2001,Rix2011,Beeks2023GrowthCaF2}.

The O-center width is then
\begin{equation}
\fwhmouter=63.7~\mathrm{kHz}\times
\frac{n_{\mathrm{bg}}+(\tilde q_{\mathrm{eff}}/e)\,n_{\mathrm{Th}}}{10^{18}~\mathrm{cm^{-3}}} \, .
\label{eq:widthlaw}
\end{equation}
 The background sets a floor
independent of $n_{\mathrm{Th}}$, and thorium adds a term that grows with it.

Two O-center widths have been
published, for crystals whose thorium densities differ by a factor of $83$ but
whose widths differ by only $24\%$~\cite{Huang2026NuclearClock,
Morawetz2026CwAbsorption}.  The crystals also differ in growth, processing and
O-center fraction, so we do not assume a common background.
{For a width-scale comparison we equate the reported single-feature FWHM
to $\fwhmouter$.  Convolution with an independent
scalar, instrumental, or homogeneous broadening fills the central zero
(Fig.~\ref{fig:lineshape}).  For the analytic profile the two maxima merge at
a Gaussian standard deviation $0.179\,\nuS$ or a Lorentzian HWHM
$0.281\,\nuS$.  Within this convolution model the added
broadening also widens the feature, so the following comparison bounds
$\gammaU$ from above.}

The X2 segment~\cite{Morawetz2026CwAbsorption} of \ThCaF has
$n_{\mathrm{Th}}=6.6(5)\times10^{18}~\mathrm{cm^{-3}}$ and an O-center width of
$1.1(1)~\mathrm{MHz}$ with an O-center fraction
$f_{\mathrm O}=0.35$~\cite{Hiraki2025-caf2-site-resolved}, consistent with the
integrated (amplitude times width) absorption of the O and D centers in Ref.~\cite{Morawetz2026CwAbsorption}.
With this fraction Eq.~\eqref{eq:qeffsplit} gives
$\tilde q_{\mathrm{eff}}=2.1$--$2.4\,e$, and the thorium term
of Eq.~\eqref{eq:widthlaw}  alone predicts an O-center width
$\fwhmouter=0.86$--$1.0~\mathrm{MHz}$ for $A_{\mathrm{eff}}=180$ and
$1.0$--$1.2~\mathrm{MHz}$ for $A_{\mathrm{eff}}=230$.
The thorium-linked contribution is therefore comparable to the measured
width.

{The X2 line was recorded with a cavity-stabilized VUV laser whose
linewidth was estimated at $\lesssim10~\mathrm{kHz}$;  the measured $91(2)~\mathrm{kHz}$ D-center line width gives
an upper bound~\cite{Morawetz2026CwAbsorption}.  At
$\nuS=1.70~\mathrm{MHz}$ the analytic maxima lie $0.44~\mathrm{MHz}$ apart,
around a dip $0.21~\mathrm{MHz}$ wide.  A $91~\mathrm{kHz}$ Lorentzian kernel
leaves the analytic dip $71\%$ deep and the anisotropic sampled dips
$62$--$65\%$ deep.  Within the independent-convolution model, reducing the
contrast below $10\%$ while retaining the measured $1.1~\mathrm{MHz}$ width
requires a Gaussian or Lorentzian kernel of
$\approx 0.28$--$0.36~\mathrm{MHz}$ FWHM.  The
intrinsic EFG outer FWHM is then $0.66$--$0.75~\mathrm{MHz}$ for the analytic
profile.  An additional tensor EFG cannot be
represented by this kernel and must instead be included before diagonalization.

The S1 crystal~\cite{Huang2026NuclearClock} has
$n_{\mathrm{Th}}=8\times10^{16}~\mathrm{cm^{-3}}$ and an O-center width of
$0.89(4)~\mathrm{MHz}$, measured in fluorescence with a sub-Hz-width laser.
{It is probed within a region that contains growth bubbles, whose strain
can modify the central suppression. A finite central
value requires effectively 1D total EFG statistics or an
additional distribution of isomer (chemical) shifts.  Whether the bubble environment
realizes either condition depends on its geometry and optical sampling.
About $80\%$ of its \thor occupies O centers~\cite{Huang2026NuclearClock},
yet Eq.~\eqref{eq:qeffsplit} still gives a thorium term of only
$\approx 20~\mathrm{kHz}$.  {The S1 width then bounds the point-defect
background of the probed region, $n_{\mathrm{bg}}\le1.4\times10^{19}~\mathrm{cm^{-3}}$,
or about $230~\mathrm{ppm}$ of the fluorine sublattice as dissociated anion
Frenkel pairs.}  Charge neutrality within the background fixes its
positive-charge density at {$n_{\mathrm{bg}}/2.4\le6\times10^{18}~\mathrm{cm^{-3}}$}
almost independently of its composition: a Frenkel pair and, for
a modest misfit, a trivalent cation with a remote $\mathrm F_i'$  carry about
$2.4\,e$ of EFG source charge.  If the positive defects are fluorine
vacancies, at most about {$120~\mathrm{ppm}$} of the fluorine sites are empty,
a  measurable quantity~\cite{Bollmann1972ConcentrationMobility,Sils2007OxygenDefectsCaF2}.

As to other broadening contributions, (i) magnetic-dipolar coupling of  the \ce{^19F} nuclear
spins to \thor spins contributes $\sim 1.0~\mathrm{kHz}$ and (ii) the natural radiative linewidth is negligible.
{(iii) A remote defect can also shift the transition as a whole through
the isomer (chemical) shift: its strain and, if it is charged, its field change
the thorium-projected valence-band occupations that set this
shift~\cite{perera2025-isomer-shift-Th229}. Measured width of the D-center line restrict it to $ \lesssim  100 \, \mathrm{kHz}$.
A quantitative, site-specific
electronic-structure treatment remains to be developed.}

 We conclude that the non-observation of the central spectral dip in Refs.~\cite{Huang2026NuclearClock,
 Morawetz2026CwAbsorption} is likely due to the presence of spatially
 extended defects in \ThCaF.
The materials literature on \ce{CaF2} provides multiple candidates: axial growth bubbles, fluoride-deficiency-driven
Ca colloids and defect aggregates, irradiation-induced dislocation loops or
voids, aligned dislocation or subgrain walls, seed--melt and growth-sector
interfaces, and smooth composition or residual-stress gradients
\cite{Beeks2023GrowthCaF2,Beeks2024a,Rix2011,
Beeks2022Dissertation,JohnsonChadderton1983AnionVoidage,
Huang2026NuclearClock,ToscaniDeCol2026FeedbackLoop}.  Their presence can introduce isomer shift broadening removing the predicted dip.
For the EFG-induced broadening,  the dip would be removed if such a structure constrained the EFG
 to one fixed tensor direction, i.e. for a uniaxial growth
gradient, a planar interface, or an aligned dislocation array.
 If spectroscopically resolved in perfectly-grown crystals,
the steep inner walls of the central dip can be used for nuclear clock operation.

\begin{acknowledgments}
I am grateful to V.~Veryazov, M.~Krośnicki, U.~Perera, B.~Katz, E.~Hudson, R. Ellwel, and Sh.~Kolkowitz  for discussions
and D.~Rehn for the relaxed defect geometries from Ref.~\cite{Rehn2026DefectAssignment} used in the
size-misfit analysis.  {This work was
supported in part by the U.S. National Science Foundation under Grant
No.~PHY-2513134.}  Artificial-intelligence assistance was used for code
development; all results were verified independently.
\end{acknowledgments}

\textit{Data availability.---}The data that support the findings of this
article are not publicly available.  The data are available from the author
upon reasonable request.

\end{document}